\documentclass[a4paper]{jpconf}
\usepackage{graphicx}
\usepackage{iopams}
\begin{document}
\title{The Tamm-Dancoff Approximation as the boson limit of the Richardson-Gaudin equations for pairing}

\author{Stijn De Baerdemacker}

\address{Ghent University, Department of Physics and Astronomy, Proeftuinstraat 86, 9000 Gent, Belgium\\
Department of Physics, University of Toronto, Toronto, Ontario M5S 1A7, Canada\\
Department of Physics, University of Notre Dame, Notre Dame, IN 46556-5670, USA}

\ead{stijn.debaerdemacker@ugent.be}

\begin{abstract}
A connection is made between the exact eigen states of the BCS Hamiltonian and the predictions made by the Tamm-Dancoff Approximation.  This connection is made by means of a parametrised algebra, which gives the exact quasi-spin algebra in one limit of the parameter and the Heisenberg-Weyl algebra in the other.  Using this algebra to construct the Bethe Ansatz solution of the BCS Hamiltonian, we obtain parametrised Richardson-Gaudin equations, leading to the secular equation of the Tamm-Dancoff Approximation in the bosonic limit.  An example is discussed in depth.
\end{abstract}

\section{Introduction}

The description of superconductive properties in many-particle systems such as metals \cite{bardeen:57} or atomic nuclei \cite{bohr:58} involves the process of pairing.  Particles in time-reversed single-particle states will lower their energy by forming a pair, constituting a condensate in the ground state of the system.  This ground state is very-well approximated by the Bardeen-Cooper-Schrieffer (BCS) wavefunction \cite{bardeen:57}, which is a variational solution of the BCS Hamiltonian in the grand canonical ensemble.  While this solution is essentially sufficient for the thermodynamic limit, $N\rightarrow\infty$, it is not fit for the description of finite-size systems.  This is due to the increasing relevance of the pair fluctuations with decreasing number of particles, which cannot be taken into account in the BCS approximation \cite{rowe:10}.  This is not of paramount importance for modestly small systems, such as \emph{e.g.} single-species nucleon pairing within a single major nuclear shell \cite{heyde:94}, because these can be diagonalised numerically on a modern computer with relative ease.  However, for mesoscopic systems, such as metallic nanograins \cite{vondelft:01}, or nucleons in multiple open shells, the dimension of the Hilbert space is huge, such that straightforward numerical diagonalisation becomes practically intractable.

Fortunately, alternative methods to extract the spectroscopy of the model (BCS) Hamiltonian exist.  Richardson and Sherman \cite{richardson:63,richardson:64a} have shown that the reduced BCS Hamiltonian, with a level-independent interaction, is exactly solvable by means of a Bethe Ansatz product wave function, provided the variables in the Ansatz wavefunction provide a solution of the set of non-linear Richardson-Gaudin (RG) equations.   As a result, the problem of finding the eigenvalues and eigenstates of the BCS Hamiltonian is reduced from diagonalising a matrix that scales \emph{factorial} with the size of the system to solving a set of algebraic equations that scales \emph{linear} with the system size.  However, the RG equations are highly non-linear in the variables, and proved to be very challenging to solve \cite{richardson:64b,richardson:66}, due to the occurrence of singularities at the critical interaction values.  It was only recently that this obstacle has been sufficiently removed \cite{rombouts:04} by means of a change of variables method, canceling the singularities at the critical points.  Alongside the numerical usefulness, the method also contributed to a better understanding of the critical points \cite{dominguez:06}, which finds its place among other recent studies of the structure of the eigenstates of the exactly solvable and integrable models \cite{roman:03,sambataro:07}.  For an extensive review on the subject, I would like to refer to the recent \textit{colloquium} \cite{dukelsky:04}.

The purpose of the present manuscript is to contribute to the understanding of the Bethe Ansatz eigenstates of the BCS Hamiltonian and the associated RG equations by making a connection with the Tamm-Dancoff Approximation (TDA) \cite{rowe:70}. In the next section, the BCS model and its algebraic properties will be introduced.  In section \ref{section:tdarg}, the TDA is introduced and its relation with RG is made.  Conclusions are presented in section \ref{section:conclusions}

\section{The reduced BCS model, an exactly solvable model}\label{section:bcsmodel}
Consider a fermionic system with $n$ single-particle energy levels $\varepsilon_i$ ($1\le i\le n$)  with degeneracy $\Omega_i$.  For condensed matter physics the levels are two-fold degenerate ($\Omega_i=2$), corresponding to the two projections of the electron spin.  In the context of nuclear structure physics, we can assign a half-integer total angular momentum $j_i$ with degeneracy $\Omega_i=2j_i+1$ to every level \cite{heyde:94}.  Naturally, and in order to make the discussion as general as possible, we will treat the case of condensed matter as $j_i=\frac{1}{2}$ systems.  The reduced BCS Hamiltonian for such a general system is given by
\begin{equation}\label{bcs:hamiltonian}
 \hat{H}=\sum_{i=1}^n \varepsilon_{i}\hat{n}_{i}+g\sum_{i,k=1}^n\hat{S}^\dag_{i}\hat{S}_{k},
\end{equation}
with the operators in the Hamiltonian given by
\begin{equation}\label{bcs:fermionoperators}
 \hat{n}_{i}=\sum_{m_i=-j_i}^{j_i}\hat{a}^\dag_{j_im_i}\hat{a}_{j_im_i},\quad \hat{S}_{i}^\dag=\case{1}{2}\sum_{m_i=-j_i}^{j_i}(-)^{j_i+m_i}\hat{a}^\dag_{j_im_i}\hat{a}^\dag_{j_i-m_i},\quad \hat{S}_{i}=(\hat{S}_{i}^\dag)^\dag,
\end{equation}
which are known respectively as the particle counting, pair creation and pair annihilation operator.  The underlying algebraic structure of these operators is the $su(2)$ quasi-spin algebra.  This algebra is spanned by the pair creation/annihilation operators, completed with the operator $\hat{S}^0_{i}=\frac{1}{2}\hat{n}_{i}-\frac{1}{4}\Omega_{i}$.  We obtain
\begin{equation}\label{bcs:quasispinalgebra}
 [\hat{S}^0_{i},\hat{S}^\dag_{k}]=\delta_{ik}\hat{S}^\dag_{i},\qquad[\hat{S}^0_{i},\hat{S}_{k}]=-\delta_{ik}\hat{S}_{i},\qquad[\hat{S}^\dag_{i},\hat{S}_{k}]=2\delta_{ik}\hat{S}^0_{i}.
\end{equation}
So, the spectrum generating algebra of the pairing problem is $\bigoplus_{i=1}^n su(2)_{i}$, with $su(2)_{i}$ the quasi-spin algebra for the level $i$.  The associated Hilbert space is constructed from the direct product of the physically allowed $su(2)_{i}$ representations $|s_{i},m_{s_i}\rangle$, with the quantum numbers\footnote{I will denote the quasi-spin projection quantum number $m$ of the quasi-spin $s_i$ explicitly by $m_{s_i}$ to rule out any confusion with the total angular momentum projection $m_i$ of $j_i$.} defined by the eigenvalues of the quadratic Casimir operator $\hat{\mathcal{C}}^2_{i}$ of $SU(2)_{i}$ and its $SO(2)_{i}$ projection $\hat{S}^0_{i}$
\begin{eqnarray}
 &\hat{\mathcal{C}}^2_{i}|s_{i},m_{s_i}\rangle=s_{i}(s_{i}+1)|s_{i},m_{s_i}\rangle\\{}
 &\hat{S}^0_{i}|s_{i},m_{s_i}\rangle=m_{s_i}|s_{i},m_{s_i}\rangle.
\end{eqnarray}
The physically allowed representations are those for which the following relations hold
\begin{equation}
 m_{s_i}=\case{1}{2}n_{i}-\case{1}{4}\Omega_{i},\quad s_{i}=\case{1}{4}\Omega_{i}-\case{1}{2}v_{i},
\end{equation}
with $n_{i}$ the number of particles in a level $i$ and $v_{i}$ the \textit{seniority} of the level $i$, which is the number of particles in a level $i$ not pairwise coupled \cite{talmi:93}.  Obviously, for systems with a definite total number of particles $N$, we require the extra condition $\sum_in_{i}=N$.  For simplicity, we will assume that there are no unpaired particles present in the system ($v_{i}=0,\forall i$), so the total number of pairs is given by $N/2$.

The key ingredient of the Richardson-Gaudin treatment of the pairing model is that the Hamiltonian (\ref{bcs:hamiltonian}) can be diagonalised by means of a Bethe Ansatz wavefunction \cite{richardson:64a,richardson:64b}
\begin{equation}\label{bcs:betheansatz}
 |\psi\rangle=\prod_{\alpha=1}^{N/2}\left(\sum_{i=1}^{n}\frac{S^\dag_{i}}{2\varepsilon_{i}-E_\alpha}\right)|\theta\rangle,
\end{equation}
provided the Richardson variables, given as the free parameters $E_\alpha$, are a solution of the RG equations
\begin{equation}\label{bcs:rgequations}
 1+2g\sum_{i=1}^n\frac{s_{i}}{2\varepsilon_{i}-E_\alpha}-2g\sum_{\beta\neq\alpha}^{N/2}\frac{1}{E_\beta-E_\alpha}=0,\qquad (1\le\alpha\le N/2).
\end{equation}
The energy of the eigen state is then given by $E=\sum_{\alpha=1}^{N/2}E_\alpha$. In the next section, it will be discussed how the Richardson-Gaudin treatment can be related to the Tamm-Dancoff Approximation, which is a well-studied method in many-body physics.
\section{Richardson-Gaudin and the Tamm-Dancoff Approximation}\label{section:tdarg}
\subsection{The Tamm-Dancoff Approximation}
Most commonly, the Tamm-Dancoff Approximation is used in the context of particle-hole excitations across a closed Hartree-Fock vacuum \cite{rowe:70,heyde:94}, but we can interprete it here somewhat larger as the description of elementary excitations which are also related to the Random Phase Approximation or Equations of Motion methods \cite{rowe:68}.  The basic idea behind the approximation is that we can treat all the excited states of the Hamiltonian as harmonic excitations or multiphonon states of the elementary eigen modes or phonons of the system.  In the case of the BCS Hamiltonian (\ref{bcs:hamiltonian}), the elementary excitations can be found by solving the eigen value equation for the 1-pair excitations $\hat{b}^\dag_k=\sum_i Y_{ki} \hat{S}^\dag_{i}$ 
\begin{equation}\label{tdarg:tdaeigenvalueequation}
 \hat{H}\hat{b}^\dag_k|\theta\rangle=\hbar\omega_k\hat{b}^\dag_k|\theta\rangle.
\end{equation}
Once this is solved, the excited states are approximated by the following multiphonon states
\begin{equation}
 |\psi\rangle\approx|\phi_{[\nu_k]}\rangle=\prod_{k=1}^n (\hat{b}^\dag_k)^{\nu_k}|\theta\rangle,
\end{equation}
with $\sum_k\nu_k=N/2$ and the notation $[\nu_k]$ is short-hand for the array $[\nu_1,\nu_2,\dots,\nu_n]$.  Similarly, the energy of this state is equal to
\begin{equation}\label{tdarg:excitationenergy}
 E^\textrm{TDA}_{[\nu_k]}=\sum_{k=1}^n\hbar\omega_k\nu_k.
\end{equation}

\begin{figure}[!htb]
 \begin{center}
  \includegraphics{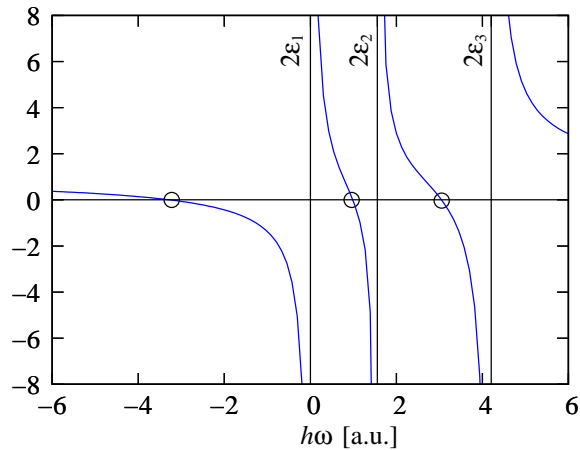}
  \caption{The TDA secular equation (eq.~(\ref{tdarg:tdaseqularequation})) for the parameters given in Table \ref{table:parameters} and an interaction strength $g=-0.5$.  The poles at the single-particle energies are accentuated by vertical lines and the 3 solutions of the equation are marked graphically by open circles.}\label{figure:tdaseqularequation}
 \end{center}
\end{figure}

Obviously, this is a gross simplification compared to the exact Bethe Ansatz wave function (\ref{bcs:betheansatz}).  However, it is worthwhile to scrutinise the differences.  For a 1-pair system, the RG and TDA approach are identical because the action of the elementary phonons $\hat{b}^\dag_k$ on the vacuum is equivalent to the Bethe Ansatz wavefunction (\ref{bcs:betheansatz}) for a single pair ($N/2=1$).  Moreover, it is well-known that the TDA eigen modes ($\hbar\omega_k$) for the reduced BCS Hamiltonian can be identified as the solutions of the secular TDA equation \cite{heyde:94,rowe:70}.
\begin{equation}\label{tdarg:tdaseqularequation}
 1+2g\sum_{i=1}^n\frac{s_{i}}{2\varepsilon_{j_i}-\hbar\omega}=0,
\end{equation}
which is identical to the RG equation for a single pair (\ref{bcs:rgequations}).  Naturally, the situation is different for multi-pair excitations.  It is instructive to continue with an example.  Consider a system of 4 pairs in 3 levels with the parameters given in Table \ref{table:parameters}.  
\begin{table}
 \caption{The parameters of a simple 3-level pairing system that serves as an example throughout the text.}\label{table:parameters}
 \begin{center}
  \begin{tabular}{l|lll}
   \br
    $i$ & $j_i$ & $\Omega_i$ & $\varepsilon_i$ \\
   \mr
   1 & $\frac{5}{2}$ &  6 & 0.0000 \\
   2 & $\frac{3}{2}$ &  4 & 0.7796 \\
   3 & $\frac{9}{2}$ & 10 & 2.1024 \\
   \br
  \end{tabular}
 \end{center}
\end{table}
The left-hand side of the secular TDA equation (\ref{tdarg:tdaseqularequation}) for the parameter set of Table \ref{table:parameters} is plotted in Figure \ref{figure:tdaseqularequation}, revealing quite a bit of information at a glance.  It is readily understood that there are as many eigen modes as there are levels and one eigen mode is structurally different from the others in the sense that it is not bound in a domain between two single-particle poles.  This solution is unbound below and is generally referred to as the \emph{collective} solution.

\begin{figure}[!htb]
 \begin{center}
 \includegraphics{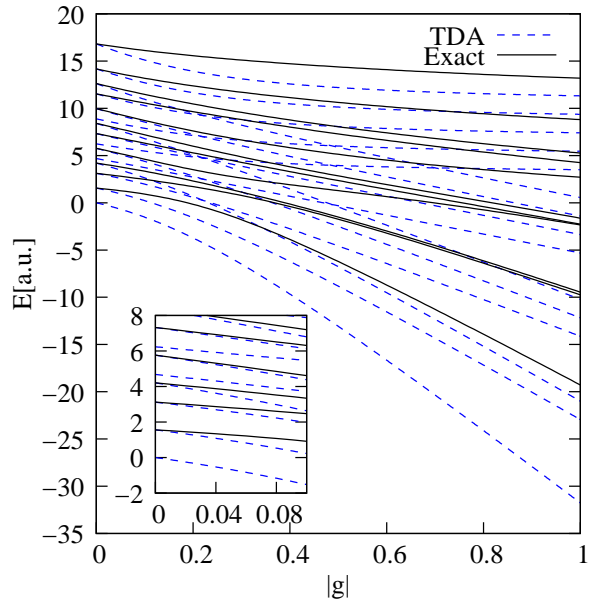}
 \caption{The exact spectrum of the BCS Hamiltonian (\ref{bcs:hamiltonian}) in (black) solid lines and the predictions of the TDA in (blue) dashed lines are plotted for a range of interaction strength $g$.  In the inset, we find a zoom of the plot for small values of $g$.}\label{figure:tdavsexact}
 \end{center}
\end{figure}

So, we gather that there are 3 eigen modes for the example under discussion.  Since we are accommodating 4 pairs in the system, we can construct a total of 15 distinct TDA multiphonon states $[\nu_1,\nu_2,\nu_3]$ with $\nu_1+\nu_2+\nu_3=4$.  Comparing this with the real dimension of the Hilbert space, which is 11, we immediately realise that TDA overestimates the total number of physically allowed states.  The question arises which ones are physical and which are not.  To investigate this question, it is instructive to plot the exact excitation spectrum of the system against the predictions made by TDA, which is done in Figure \ref{figure:tdavsexact}.
We can roughly divide the Figure in 2 different domains, \emph{i.e.} the small-interaction domain $|g|\sim0$ and the large-interaction domain $|g|\gg0$.  The former domain is enlarged and depicted in the inset of the Figure.  From this inset, it is clear which TDA states are unphysical for the small-interaction domain.  We see that for $|g|\rightarrow0$, the exact ground-state energy is given by approximately $6\varepsilon_1+2\varepsilon_2=1.5592$, whereas the TDA predicts the ground state at $8\varepsilon_1=0.0000$.  This is because the TDA can put all 4 elementary excitations in the lowest level, $[\nu_1,\nu_2,\nu_2]=[4,0,0]$, whereas this is fundamentally forbidden in the Hilbert space because of the Pauli principle.  The degeneracy of the lowest level $\Omega_1=6$ only allows for 3 pairs, so one pair needs to be promoted to the 2nd level.  As a result, we can identify $[4,0,0]$ as physically forbidden in this domain.  Further investigation reveals that $[1,3,0]$, $[0,4,0]$ and $[0,3,1]$ are also forbidden on similar grounds.  The situation is less clear for the large-interaction domain because we cannot rely on pure combinatorial arguments.  However, we can draw qualitative conclusions by noticing that the excitation spectrum tends to reorganise in small clusters around $|g|=1$.  We count clusters of 1, 2 and 3 states from the bottom up and 5 more states that have no cluster characterisation.  Inspecting the predictions by TDA, we notice a similar pattern where every cluster can be associated with a distinct number of collective phonons ($[4,0,0]$ for the ground state, $[3,1,0]$ and $[3,0,1]$ for the next cluster, etc.).  Nevertheless, we also notice that TDA systematically underestimates the excitation energy of the corresponding states.  Again, this is a manifestation of the Pauli exclusion principle, however in a different way.  The correct clustering for the lowest states reveals that TDA gives a correct idea about the collective structure of the lowest states, but TDA allows Pauli-forbidden configurations to contribute to the energy of the eigen state.  Anyhow, these qualitative arguments do not answer the question which TDA states are related to which exact eigenstates and which TDA states are not physical.  In the next subsection, I will present a construction for this particular purpose.
\subsection{An algebraic approach to go from Tamm-Dancoff to Richardson-Gaudin}
We may conclude from the previous subsection that the difference between TDA and the exact solution are due to the Pauli principle.  In algebraic phrasing, the fundamental algebra of the exact pairing model is quasi-spin $su(2)$ whereas TDA is characterised by the bosonic Heisenberg-Weyl algebra $hw(1)$, given by \cite{rowe:10}.
\begin{equation}
 [\hat{b}_k,\hat{b}^\dag_l]=\delta_{kl}\hat{1},\quad [\hat{b}_k,\hat{1}]=[\hat{b}^\dag_k,\hat{1}]=0.
\end{equation}
From a physical point of view, one can conclude that the fermion pairs $\hat{S}^\dag_i$ (eq.~(\ref{bcs:fermionoperators})) have been replaced by genuine bosons $\hat{b}_k$ in the TDA.  This substitution is rather abrupt, so it would be interesting to have a more adiabatic means to \emph{bosonise} the fermion pairs.  This can be realised by introducing the following parametrised algebra
\begin{equation}\label{tdarg:deformedalgebra}
 [\hat{S}^0_{i},\hat{S}^\dag_{k}]=\delta_{ik}\hat{S}^\dag_{i},\qquad[\hat{S}^0_{i},\hat{S}_{k}]=-\delta_{ik}\hat{S}_{i},\qquad[\hat{S}^\dag_{i},\hat{S}_{k}]=\delta_{ik}(\xi2\hat{S}^0_{i}+(\xi-1)\case{1}{2}\Omega_i\hat{1}),
\end{equation}
with $\xi$ a real-valued parameter between 0 and 1.  On the one hand, we retain the quasi-spin algebra for $\xi=1$, and on the other hand, we get the Heisenberg-Weyl algebra in the $\xi=0$ limit, if we scale the boson creation- and annihilation operators with a factor $\sqrt{2/\Omega_i}$, and interprete the operator $\hat{S}^0_{i}$ as a boson counting operator.  In the intermediate region, we can use the parameter $\xi$ as a continuous switch to transform the fermion pairs into bosons and the other way around.  This construction is closely related to the contraction of $u(2)$ to $hw(1)$, as defined by Arecchi \emph{et.~al.} \cite{arecchi:72,gilmore:08}.  The question is now whether this functionality is also transferred to the pairing system.  Remarkably, if we use the parametrised algebra (\ref{tdarg:deformedalgebra}) instead of the original quasi-spin algebra (\ref{bcs:quasispinalgebra}), we observe that the system remains exactly solvable by means of a Bethe Ansatz wavefunction, provided the free variables in the wavefunction are a solution of the following parametrised RG equations
\begin{equation}\label{tdarg:deformedrgequations}
  1+g\sum_{i=1}^n\frac{\frac{1}{2}\Omega_i-\xi v_i}{2\varepsilon_{i}-E_\alpha}-2g\xi\sum_{\beta\neq\alpha}^{N/2}\frac{1}{E_\beta-E_\alpha}=0,\qquad (1\le\alpha\le N/2).
\end{equation}
It is readily verified that this set of equations reduces to the correct corresponding expressions in the limits for $\xi$.  For $\xi=1$, we retain the set of RG equations (eq.~(\ref{bcs:rgequations})), and for $\xi=0$, we obtain $N/2$ identical copies of a (seniority zero ($v_i=0$, $\forall i$)) secular TDA equation.  This means in the latter case, that every individual pair in the Bethe Ansatz wavefunction needs to satisfy the TDA secular equation, independent from the other pairs because the coupling term between the variables $E_\alpha$ in the eqs.~(\ref{tdarg:deformedrgequations}) has disappeared at $\xi=0$.  More details on the technical aspects of these results, as well as on the nature of the algebra (\ref{tdarg:deformedalgebra}) will be given in a forthcoming paper.

\begin{figure}[!htb]
 \begin{center}
  \includegraphics{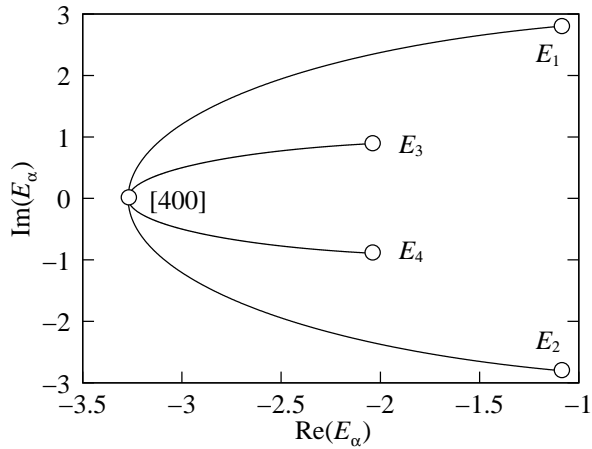}
  \caption{The $\xi$-trajectory of a solution of the parametrised RG equations (\ref{tdarg:deformedrgequations}) in the complex plane.  We start from the $[4,0,0]$ TDA solution and adiabatically go to the corresponding solution of the RG equations (\ref{bcs:rgequations}) for $\xi=1$ (denoted by $E_\alpha$, $\alpha=1\dots4$).  The parameters for the model are given in Table \ref{table:parameters} and $g=-0.5$.}\label{figure:trajectoriesrealvsimag}
 \end{center}
\end{figure}

Once we have the $\xi$-parametrised RG equations, we can investigate how the exact eigen states of the pairing problem are related to the solutions of the TDA.  For this purpose, we can start from a known solution of the pairing problem \cite{rombouts:04} and see which TDA solution corresponds to this solution, or \emph{vice versa}, we can start from a particular multiphonon solution of TDA and reconstruct the corresponding exact solution, if it exists.  This can be done in both cases by adiabatically switching off/on the $\xi$ parameter in the eqs.~(\ref{tdarg:deformedrgequations}) and solving numerically along the trajectory.  One particular result for the present example (with parameters from Table \ref{table:parameters} and $g=-0.5$) can be found in Figure \ref{figure:trajectoriesrealvsimag}.   We start from the $[4,0,0]$ multiphonon TDA solution of the system (see also Figure \ref{figure:tdaseqularequation} for the TDA eigen modes) and, by adiabatically switching on the parameter $\xi$, obtain the solution of the RG equations.  In the Figure, the $\xi$-trajectory of the solution in the complex plane is depicted.

\begin{figure}[!htb]
 \begin{center}
 \includegraphics[width=14cm]{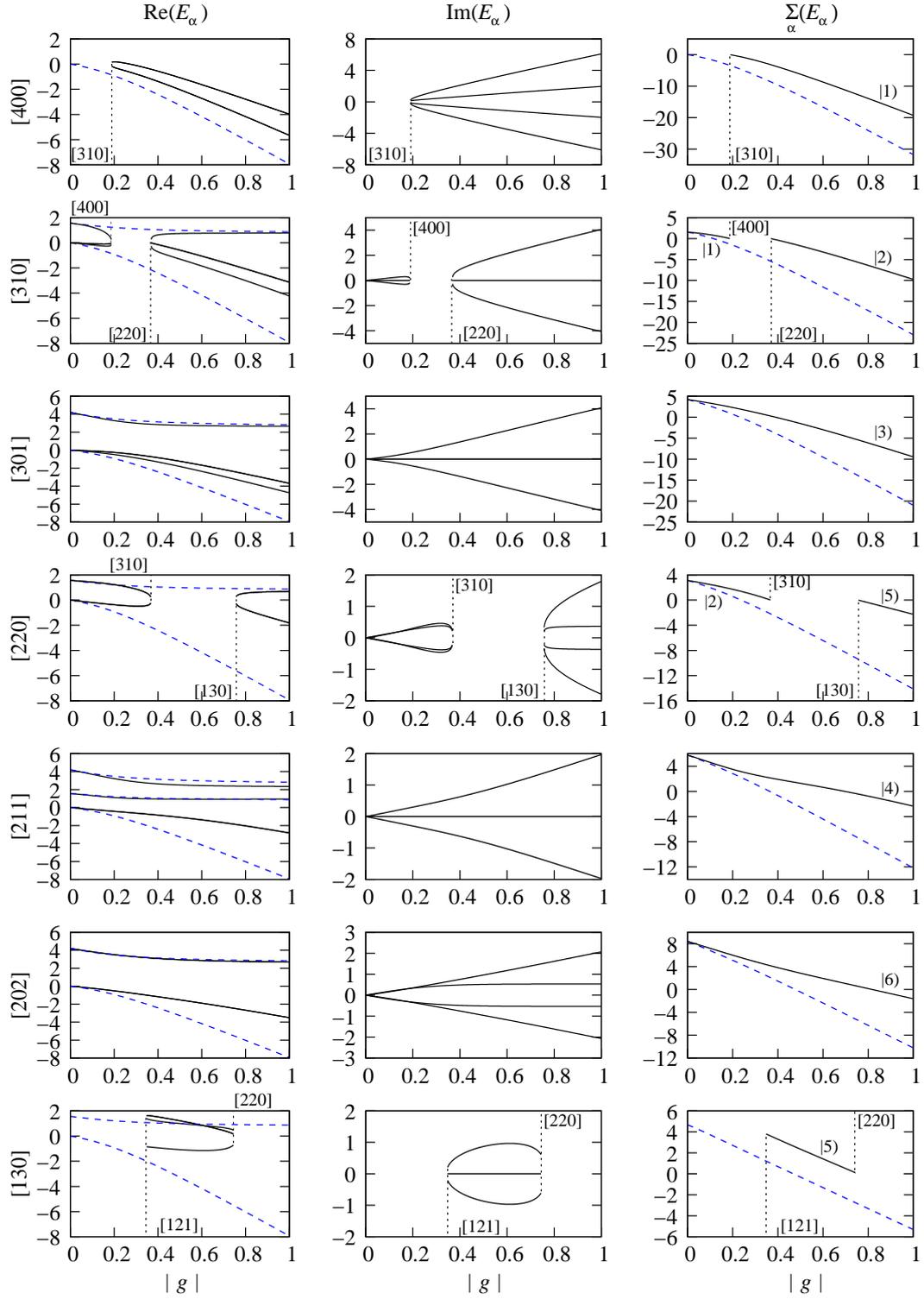}
 \caption{The TDA eigen modes and corresponding RG variables for given TDA multiphonon states as a function of the interaction $|g|$.  More information is discussed in the text.}\label{figure:tda2rg-allthestates-part1}
 \end{center}
\end{figure}

\begin{figure}[!htb]
  \begin{center}
  \includegraphics[width=14cm]{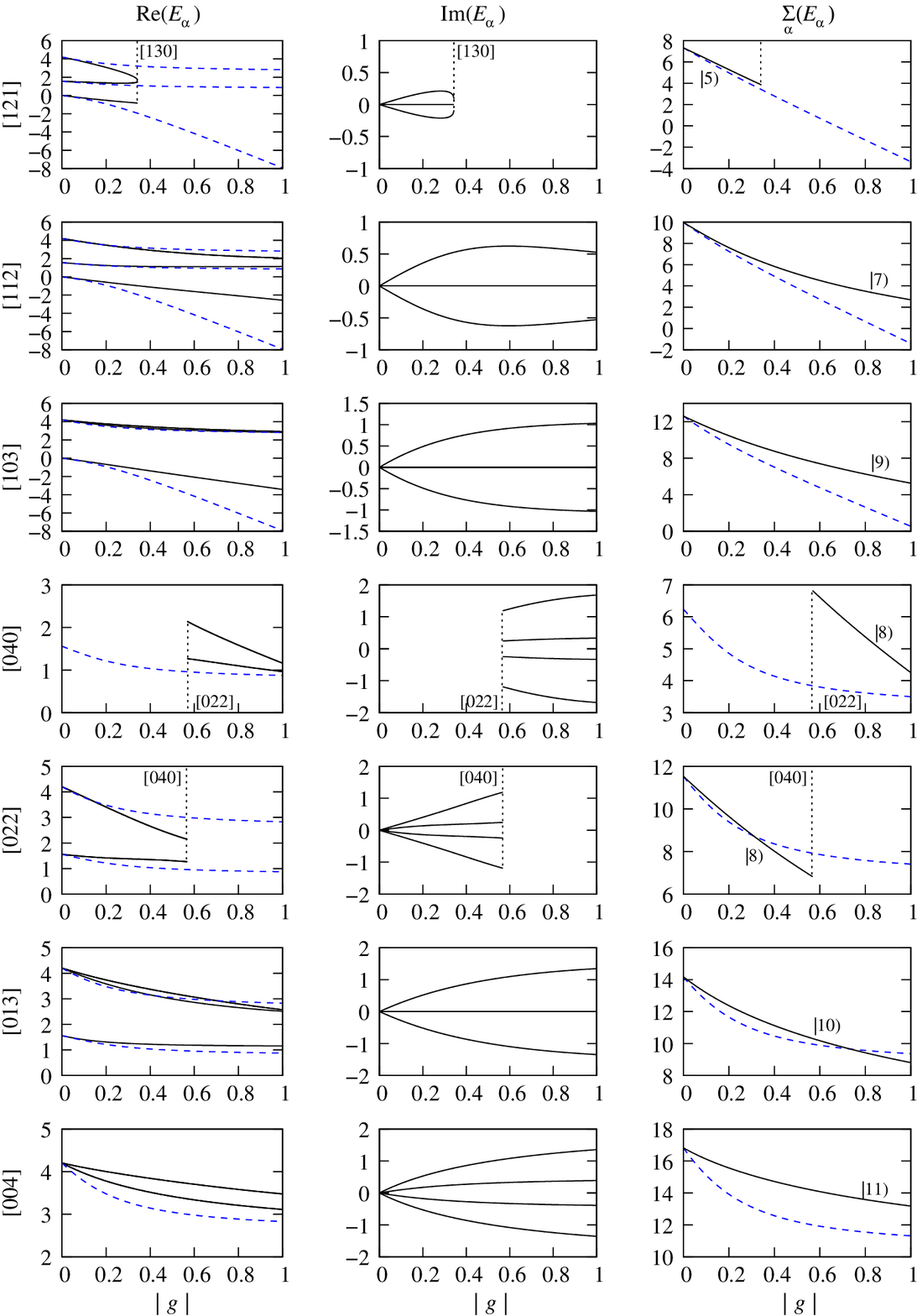}
 \caption{The TDA eigen modes and corresponding RG variables for given TDA multiphonon states as a function of the interaction $|g|$.  More information is discussed in the text.}\label{figure:tda2rg-allthestates-part2}
  \end{center}
\end{figure}

Finally, we can use this method to make a one-to-one connection between the exact eigen states and the predicted states by TDA, plotted in Figure \ref{figure:tdavsexact}.  This is done extensively in Figure \ref{figure:tda2rg-allthestates-part1} and \ref{figure:tda2rg-allthestates-part2}.  Each row of panels corresponds to a particular TDA state, denoted by $[\nu_1,\nu_2,\nu_3]$.  In the first panel, the energy of the elementary TDA eigen modes (blue dashed line) and the real part of the corresponding RG variables (black full line) are plotted as a function of the interaction strength $|g|$.  The second panel shows the imaginary part of the RG variables, and the third panel shows the total energy of the exact eigenstate $\sum_{\alpha=1}^4E_\alpha$ (black full line) as well as the TDA prediction of the total energy (eq. (\ref{tdarg:excitationenergy})) (blue dashed line) as a function of $|g|$.   For every TDA state $[\nu_1,\nu_2,\nu_3]$, the interaction strengths $g$ is scanned and the corresponding RG variables are given if they exist.  In some cases, such as \emph{e.g.} for the low-interaction domain of $[4,0,0]$ (first row of Figure \ref{figure:tda2rg-allthestates-part1}), the TDA state did not produce an exact solution and it is only for values larger than $|g|\sim0.190$ that it leads to an exact state (the ground state).  It is readily seen from the second row that the ground state below this value $|g|\sim0.190$ is connected to the $[3,1,0]$ TDA state.  Dotted lines are added to the figures to guide the eye at those transitional points where the TDA origin of the exact states changes from one state to another.  In addition, a label $|i)$ is added in the third panel of every row to denote the corresponding exact state.  We can draw multiple conclusions from these figures.  It can be observed that the TDA states $[4,0,0]$, $[1,3,0]$, $[0,3,1]$ (not depicted because it does not lead to any exact solution for $|g|\in[0,1]$) and $[0,4,0]$ do not correspond to an exact solution in the small-interaction domain, which is consistent with what we derived earlier.  Some states are clearly connected to a given state for the whole domain of $|g|\in[0,1]$ (such as \emph{e.g.}\, $[3,0,1]$ with the 3rd excited state), whereas other TDA states have no corresponding exact solution over the domain (such as \emph{e.g.}\, $[0,3,1]$).  It can also occur that one TDA state can connect to different exact states, however for different domains of $|g|$ (such as \emph{e.g.}\, $[3,1,0]$, which gives rise to the ground state for small $|g|$ as well as to the 2nd excited state for larger $|g|$).  In addition, we can verify that the observed clustering around $|g|\sim1$ is indeed related to the number of collective TDA eigen modes $\nu_1$, consistent with the earlier observations.

\section{Conclusions}\label{section:conclusions}
In conclusion, an algebraic technique to connect the exact eigen states of the BCS Hamiltonian with the predictions of Tamm-Dancoff Approximation is presented.  The connection is made via a parametrised algebra, giving rise to the quasi-spin $su(2)$ algebra and the Heisenberg-Weyl $hw(1)$ in the two limits of the parameter.  Using this parametrised algebra for the Bethe Ansatz wavefunction, we obtain the corresponding parametrised Richardson-Gaudin equations, which reduce to the exact Richardson-Gaudin equations in the quasi-spin limit and the Tamm-Dancoff secular equation in the Heisenberg-Weyl limit.  These parametrised RG equations allow us to make a clear-cut connection between the exact states of the BCS Hamiltonian and the TDA predictions.

\ack

It is a pleasure to acknowledge P.~Van~Isacker, S.~Rombouts, D.~J.~Rowe and D.~Van~Neck for many stimulating and interesting discussion sessions.  I would also like to thank V.~Hellemans, K.~Heyde and B.~Verstichel for discussions and interesting suggestions.  Financial support comes from the "FWO-Vlaanderen" as well as 2 scholarships for a "Long stay abroad'' to support my stay at the University of Toronto and the University of Notre Dame.

\section*{References}

\bibliography{debaerdemacker-tda2rg-arxiv}
\bibliographystyle{iopart-num}

\end{document}